\documentclass{llncs}
\usepackage{psfig,subfigure}
\usepackage{enumerate,latexsym,llncsdoc}
\renewcommand\footnotemark{}
\newcommand{\define}{\stackrel{\triangle}{=}}
\begin{document}
\title{{\Huge A Game Theoretic Analysis of Collaboration in 
Wikipedia}}
\author{S.~Anand$^*$, O.~Arazy$^\dagger$, N.~B.~Mandayam$^*$ and O.~Nov$^\ddagger$}
\institute{$^*$ Wireless Information  Networks Laboratory (WINLAB), Rutgers University\\ 
E-mail: {\tt \{anands72,narayan\}@winlab.rutgers.edu}\\
$^\dagger$ Alberta School of Business, Edmonton\\
 E-mail: {\tt ofer.arazy@ualberta.ca}\\
$^\ddagger$ Department of Technology Management and  Innovation,\\ 
Polytechnic Institute of New York University\\ 
E-mail: {\tt on272@nyu.edu}\thanks{Proceedings of the $4^{th}$ International 
Conference on Decision and Game Theory for Security (GameSec 2013),  
LNCS 8252, pp. 29-44, Springer 2013}} 
\maketitle
\begin{abstract}
Peer production projects such as Wikipedia or open-source 
software development allow volunteers to collectively 
create knowledge-based products. The inclusive nature 
of such projects poses difficult challenges for ensuring 
trustworthiness and combating vandalism. Prior studies in 
the area deal with descriptive aspects of peer production, 
failing to capture the idea that while contributors 
collaborate, they also compete for status in the community 
and for imposing their views on the product. In this paper, 
we investigate collaborative authoring in Wikipedia, where
contributors append and overwrite previous contributions to 
a page. We assume that a contributor's goal is to maximize 
ownership of content sections, such that content owned 
(i.e. originated) by her survived the most recent revision 
of the page. We model contributors' interactions to increase 
their content ownership as a non-cooperative game, where a 
player's utility is associated with content owned and cost 
is a function of effort expended. Our results capture 
several real-life aspects of contributors interactions 
within peer-production projects. Namely, we show that at 
the Nash equilibrium there is an inverse relationship 
between the effort required  to make a contribution and 
the survival of a contributor's content. In other words, 
majority of the content that survives is necessarily 
contributed by experts who expend relatively less effort 
than non-experts. An empirical analysis of Wikipedia 
articles provides support for our model's predictions. 
Implications for research and practice are discussed in 
the context of trustworthy collaboration as well as vandalism.
\end{abstract}
{\em Index Terms-}{\footnotesize Peer production, Wikipedia, 
collaboration, non-cooperative game, trustworthy collaboration,
vandalism}
\section{Introduction}\label{sec:intro}
Recent years have seen the emergence of a web-based peer-production 
model for collaborative work, whereby large numbers of individuals 
co-create knowledge-based goods, such as Wikipedia, and open source 
software \cite{novkuk}, \cite{oferodedMIS}, \cite{wolf},
\cite{benkler}, \cite{MSmalla}, \cite{LeeOS}, \cite{hippel}. 
Increasingly, individuals, companies, government agencies and other 
organizations rely on peer-produced products, stressing on the need to ensure trust worthiness
of collaboration (e.g., deter vandalism) as well as the quality of end products. 

Our focus in this study is Wikipedia, probably the most prominent 
example of peer-production. Wikipedia has become one of the most 
popular information sources on the web, and the quality of Wikipedia 
articles has been the topic of recent public debates. Wikipedia is 
based on wiki technology. Wiki is a web-based collaborative authoring 
tool that allows contributors to add new content, append existing 
content, or delete and overwrite prior contributions. Wikis track 
the history of revisions – similarly to version control systems 
used in software development – allowing users to revert a wiki page 
to an earlier revision \nocite{teece}\nocite{martin}\nocite{peng}\nocite{winter}
\cite{leuf}, \cite{wagner2004}, \cite{wagner2007}.

Peer production projects face a key tension between inclusiveness 
and quality assurance. While such projects need to draw in a large 
group of contributors in order to leverage ``the wisdom of the crowd,'' 
there is also requirement for accountability, security, and quality control 
\cite{forte}, \cite{kittur}, \cite{towne}. 
Quality assurance measures are necessary not only in cases of 
vandalism; conflicts between contributors could also result from 
competition. For example, contributors to Wikipedia may wrestle 
to impose their own viewpoints on an article – especially for 
controversial topics – or attempt to dominate subsets of the 
peer-produced product. Another example is when contributors 
seeking status within the community compete to make the largest 
contribution, and in the process overwrite others' previous 
contributions. The result of such competitions is often ``edit wars'' 
where articles are changed back-and-forth between the same contributors.

Prior studies investigating an individual's motivation for 
contributing content to Wikipedia have identified a large number 
of motives driving participation\cite{oferodedMIS},\cite{krogh}, 
including motives that are 
competitive in nature, such as ego, reputation enhancement, 
and the expression of one's opinions \cite{krogh}, \cite{MISQuarterly}. 
However, studies investigating individuals did not consider 
the competitive dynamics emerging from motives such as reputation. 
Research into group interactions at Wikipedia, have tended to 
emphasize the collaborative (rather than competitive nature 
of interactions) \cite{oferodedMIS}. Other studies  investigated threats 
to security and trustworthiness resulting 
from malicious attacks (i.e. vandalism)\cite{krogh} and the organizational 
mechanisms used by Wikipedia to combat such attacks \cite{MISQuarterly}; 
yet these studies do not consider threats resulting from benevolent 
contributors. A relevant strand of the literature has looked at
conflicts of opinions between contributors \cite{oferjasist}, \cite{oferodedMIS}. 
However, the focus is on the result of these conflicts on content 
quality rather than the competitive mechanisms driving them. 
In summary, while peer-production projects, and in particular 
Wikipedia, have attracted significant attention within the research 
community, to the best of our knowledge, the competitive dynamics 
have not been investigated.

In order to better understand collaboration in Wikipedia and 
capture the competitive nature of interactions, we turn to game 
theory. Our underlying assumption is that a contributor's goal 
is to maximize ownership of content sections, such that 
content ``owned'' (i.e. originated) by that user survived 
the most recent revision of the page. We model contributors' 
interactions to increase their content ownership, as a 
non-cooperative game. A contributor's motivation for trying to 
maximize her ownership within a certain topical page 
could be based on the need to express one's views or to 
increase her status in the community; and competition 
could be the result of battles between opposing viewpoints 
(e.g. vandals and those seeking to ensure trustworthiness 
of content) or consequences of power struggles. 
The utility of each contributor in the non-cooperative game
is the ownership in the page, defined as the fraction of content
owned by the contributor in the page. Each contributor suffers
a cost of contribution which is the effort expended towards
making the contribution. The objective is then to
determine the optimal strategies, i. e., the optimal number of
contributions made by each contributor, so that her \emph{net utility}
is maximized. Here, the net utility is the difference between the
utility (a measure of the ownership) and the cost (a measure of the 
effort expended). The optimal
strategies are determined by determining the Nash equilibrium
of the non-cooperative game that models the interactions between
the contributors. We determine the conditions under which
the Nash equilibrium of the game can be achieved and find
its implications on the contributors' expertise levels on the
topic. We report of an empirical analysis of Wikipedia that 
validates the model's predictions.
The  key results brought forth by our analysis include
\begin{itemize}   
\item The ownership of contributors increases with the
decreasing levels of effort expended by the contributor
on the topic. 
\item Contributors expending 
equal amount of effort end up with equal ownership.
\end{itemize}

The rest of the paper is organized as follows. The non-cooperative
game that models the interactions between contributors is
described in Section \ref{sec:ncgame}. We then use Wikipedia 
data to validate the  modeling in Section \ref{sec:results}. 
We then discuss in Section \ref{sec:twcvand} the relevance of 
our analysis and modeling to trust worthy collaboration and 
vandalism. Conclusions are drawn in Section \ref{sec:concl} 
along with pointers to future directions.
\section{User Contribution as a Non-Cooperative Game}\label{sec:ncgame}
We model the interactions of the $N$ content
contributors to a page (i.e., users) as a non-cooperative game. 
The strategy set for each contributor is the amount and type
of contribution she makes. Table
\ref{tbl:notation} describes the notations used in our
analysis. 
and their descriptions.
\begin{table}[]
\begin{center}
\caption{\label{tbl:notation} Variables used
in the analysis in this paper.}
\begin{tabular}{|l|l|}
\hline
Notation/Variable & Description \\
\hline
$N$ & Number of users or content contributors\\
$x_i$ & The amount of content owned by the $i^{th}$ user\\
$\beta_i$ & Effort expended by user $i$ to make unit contribution\\
$u_i$ & The fractional ownership held by the $i^{th}$ user\\
$n_i$ & Net utility of contributor $i$\\
$\mathbf{1}$ & The all-one vector\\
$\mathbf{I}$ & The identity matrix\\
\hline
\end{tabular} 
\end{center}
\end{table}

Let $x_i$ represent the content owned by the $i^{th}$ user 
in the current version of the page. We define the utility, $u_i$, as
the fraction of content owned by the $i^{th}$ contributor, 
and is given by
\begin{eqnarray}\label{eqn:utilityi}
u_i= \frac{x_i}{\sum_{j=1}^N x_j}.
\end{eqnarray}
The objective of contributor $i$ is to determine the
optimal $x_i$ so that $u_i$ is maximum. 

It is observed from (\ref{eqn:utilityi})
that the optimal $x_i$ that maximizes $u_i$ is
$x_i=\infty$. This is because the utility function is
an increasing function of $x_i$. Intuitively, this result
occurs because every time the $i^{th}$ user makes a contribution,
his/her ownership increases. However this results in 
reduction in the ownership of other contributors, to counter
which, they attempt to make additional contributions 
(by increasing their respective $x_k$'s). This, in turn,
reduces the ownership of contributor $i$, thereby causing
him/her to further increase $x_i$ to increase ownership. 
This process continues ad infinitum resulting in 
$x_i\rightarrow\infty$,
$\forall$ $i$. This degenerate scenario can be mitigated
as follows.

Suppose the $i^{th}$ contributor expends an effort, 
$\beta_i$, to make a unit contribution. For instance,
$\beta_i$ can be the cost incurred by the $i^{th}$ user, 
in terms of time and effort spent in learning the topic 
and in posting content on a Wiki page.
Therefore, the  $i^{th}$ contributor expends 
a total effort $\beta_ix_i$, to achieve
$x_i$ amount of content ownership in the page. The net utility experienced
by the $i^{th}$ contributor, $n_i$, can be written as the difference
between utility of contributor $i$, given by (\ref{eqn:utilityi})
and the total effort expended by contributor $i$, i. e.,
\begin{eqnarray}\label{eqn:netutilityi}
n_i=u_i-\beta_ix_i=
\frac{x_i}{\sum_{j=1}^N x_j}
-\beta_ix_i.
\end{eqnarray}
It is observed that the net utility obtained
by the $i^{th}$ contributor not only depends on the
strategy of the $i^{th}$ contributor ({\it i.e.}, $x_i$),
but also on the strategies of all the other contributors
({\it i.e.}, $x_j$, $j\neq i$). This results
in the non-cooperative game of complete
information \cite{neumannmorgen} between 
the contributors. The optimal $x_i$,
$\forall$ $i$ (termed as $x_i^*$), which is
determined by maximizing $n_i$ in (\ref{eqn:netutilityi}),
is then the Nash equilibrium of the non-cooperative game
where no contributor can make a unilateral change.

Applying the first order necessary condition to
(\ref{eqn:netutilityi}), $x_i^*$
is obtained as the solution to
\begin{eqnarray}\label{eqn:firstderivative}
\begin{array}{cc}
\left .
\frac{\partial n_i}{\partial x_i}
\right |_{x_i=x_i^*}=
\frac{\sum_{k = 1\atop k \neq i}^{N} x_k^*}
{\left (\sum_{j = 1}^{N} x_j^*\right )^2} - \beta_i=0,
& \forall i
\end{array}
\end{eqnarray}
subject to the constraints
$x_i^*\geq 0$, $\forall$ $i$.
From (\ref{eqn:firstderivative}), we obtain
$\frac{\partial^2 n_i}{\partial x_i^2} = - \frac{2
\sum_{k = 1\atop k \neq i}^{N} x_k}
{\left (\sum_{j = 1}^{N} x_j\right )^3} < 0$,
$\forall$ $i$, when $x_i\geq 0$.
Thus, $n_i$ is a concave function of $x_i$
and $x_i^*$, which solves (\ref{eqn:firstderivative})
subject to $x_i^* \geq 0$, $\forall$ $i$,
is a local as well as a global maximum point.
In other words, according to \cite{nash},
\emph{the non-cooperative game
has a unique Nash equilibrium},
$\mathbf{x}^*=
\left [
\begin{array}{cccc}
x_1^* & x_2^* & \cdots & x_N^*
\end{array}
\right ]^T$, obtained by numerically 
solving the system of $N$
non-linear equations specified by (\ref{eqn:firstderivative}).
However, to study the effect of the effort levels
($\beta_i$'s) on the strategies of the contributors, it is desirable to obtain an
expression that relates the vectors, $\mathbf{x}^*$,
$\mathbf{x}=\left [x_i\right ]_{1\leq i\leq N}$
and $\mbox{\boldmath $\beta$}=\left [\beta_i\right ]_{1\leq i\leq N}$.

Re-writing (\ref{eqn:firstderivative}), 
\begin{eqnarray}\label{eqn:omegai}
\begin{array}{cc}
\left (\sum_{j=1}^N x_j^*\right )^2
-\alpha_i\sum_{j=1\atop j\neq i}^Nx_j^*=0,
& \forall N,
\end{array}
\end{eqnarray}
where $\alpha_i\define 1/\beta_i$.
Eqn. (\ref{eqn:omegai} ) can be written as
\begin{eqnarray} \label{eqn:modifiedmatrixform}
\left (\mathbf{x}^*\right )^T
\mathbf{1}\mathbf{1}^T
\mathbf{x}^*\mathbf{1}-
\mathbf{D}_{\alpha}\left (\mathbf{1}\mathbf{1}^T-\mathbf{I}\right )
\mathbf{x}^*=\mathbf{0},
\end{eqnarray}
where $(.)^T$ represents the transpose of
a vector or a matrix, $\mathbf{D}_\alpha$ is the diagonal
matrix $\mathbf{diag}\left (\alpha_1,
\alpha_2,\cdots,\alpha_N\right )$,
$\mathbf{1}$ is the column vector in
which all entries are one, $\mathbf{0}$ is the column
vector in which all entries are zero and $\mathbf{I}$
is the identity matrix. 

It can be easily verified the vectors,
$\mathbf{y}_1=\left [
\begin{array}{cccccc}
\frac{1}{\sqrt{N}} &
\frac{1}{\sqrt{N}} &
\frac{1}{\sqrt{N}} &
\frac{1}{\sqrt{N}} &
\cdots &
\frac{1}{\sqrt{N}}
\end{array}
\right ]^T$
and for $j=2$, $3$, $\cdots$, $N$,
$\mathbf{y}_j=
\left [
\begin{array}{cccccc}
y_{1j} & y_{2j} & y_{3j} & \cdots & y_{(N-1)j} & y_{Nj}
\end{array}
\right ]^T$,
where
\begin{eqnarray}\label{eqn:yj}
y_{kj}=\left \{
\begin{array}{cc}
\frac{1}{\sqrt{j(j-1)}} & k<j\\
-\frac{j-1}{\sqrt{j(j-1)}} & k=j\\
0 & k>j,
\end{array}
\right .
\end{eqnarray}
form a set of orthonormal eigen vectors
to the matrix, $\mathbf{1}\mathbf{1}^T$.
The eigen value corresponding to $\mathbf{y}_1$
is $N$ and those corresponding to $\mathbf{y}_2,\cdots,\mathbf{y}_N$
are $0$s.
Let $\mathbf{P}=\left [
\mathbf{y}_1\vert\mathbf{y}_2\vert\cdots\vert\mathbf{y}_N\right ]$.
Then, $\mathbf{P}$ is an orthogonal matrix and by orthogonality
transformation,
\begin{eqnarray}\label{eqn:orthogonality}
\mathbf{P}^T\mathbf{1}\mathbf{1}^T\mathbf{P}=\mathbf{D}=
\mbox{diag}\left (N, 0, 0, \cdots, 0\right ).
\end{eqnarray}
Let $\mathbf{z}=\left [
\begin{array}{cccccc}
z_1 & z_2 & z_3 & \cdots & z_{N-1} & z_N
\end{array}
\right ]^T$.
Since the eigen vectors of a matrix form a basis for the
$N-$dimensional sub-space \cite{meyerbook},
the vector, $\mathbf{x}^*$, can be written as
$\mathbf{x}^*=\mathbf{P}\mathbf{z}$. 
A similar expression has been solved in \cite{techreportpricing} 
in the context of pricing in wireless networks and we outline 
here  the key steps to determine the optimal $\mathbf{x}^*$.
\begin{itemize}
\item Using  $\mathbf{x}^*=\mathbf{P}\mathbf{z}$ in (\ref{eqn:modifiedmatrixform})
and (\ref{eqn:orthogonality}), we obtain
\begin{eqnarray}\label{eqn:nonlineZ}
\mathbf{z}^T\mathbf{D}\mathbf{z}\mathbf{1}
-\mathbf{D}_\alpha\left (\mathbf{1}\mathbf{1}^T-\mathbf{I}\right)
\mathbf{P}\mathbf{z}=\mathbf{0}.
\end{eqnarray}
 \item The above is a set of non-linear equations in
$\mathbf{z}$, in which the $k^{th}$ equation depends
on $z_1$ and $z_j$, $k\leq j\leq N$. Solving the
non-linear equations by backward substitution
\cite{meyerbook}, $z_k$, $2\leq k\leq N$ can be written in
terms of $z_1$ as 
\begin{eqnarray}\label{eqn:xkx1}
\frac{z_k}{\sqrt{k(k-1)}}=
\frac{Nz_1^2}{k(k-1)}\left [
\frac{k}{\alpha_k}+\sum_{j=k+1}^N\frac{1}{\alpha_j}\right ]
-\frac{z_1}{\sqrt{N}}\frac{N(N-1)}{k(k-1)}.
\end{eqnarray}
\item Using (\ref{eqn:xkx1}) to replace all $z_k$'s 
in terms of $z_1$ in the set of non-linear equations 
in (\ref{eqn:nonlineZ}), $z_1$ can be obtained as
\begin{eqnarray} \label{eqn:expz1}
z_1=\frac{N-1}{\sqrt{N}}
\frac{1}{G},
\end{eqnarray}
where
\begin{eqnarray}\label{eqn:defnG}
G\define\sum_{j=1}^N\frac{1}{\alpha_j}.
\end{eqnarray}
\item Combining (\ref{eqn:xkx1}) and (\ref{eqn:expz1}),
\begin{eqnarray}\label{eqn:finalexpressionxk}
\begin{array}{cc}
\frac{z_k}{\sqrt{k(k-1)}}=
\frac{(N-1)^2}{k(k-1)}G^{-1}
\left [G^{-1}\left (\frac{k}{\alpha_k}+
\sum_{j=k+1}^N\frac{1}{\alpha_j}\right )-1\right ] &
2\leq k\leq N.
\end{array}
\end{eqnarray}
\item  Using the facts
$\mathbf{x}^*=\mathbf{P}\mathbf{z}$,
and $\alpha_i=\frac{1}{\beta_i}$ in
(\ref{eqn:expz1}) and (\ref{eqn:finalexpressionxk}), 
the unique Nash equilibrium of the non-cooperative game
can be obtained as
\begin{eqnarray}\label{eqn:bistar}
x_i^*=\frac{\sum_{j=1}^N\beta_j-(N-1)\beta_i}
{\left (\sum_{j=1}^N\beta_j\right )^2}.
\end{eqnarray}
\end{itemize}
Note that the unique Nash equilibrium $\mathbf{x}^*$,
is feasible, {\it i.e.}, $x_i^* >0$, $\forall$ $i$
if and only if
\begin{eqnarray}\label{eqn:conditionnash}
(N-1)\beta_i< \sum_{j=1}^N\beta_j.
\end{eqnarray}
The utility (ownership) of contributor $i$ at the Nash equilibrium, $u_i^*$, 
can then be obtained
from (\ref{eqn:utilityi}) and (\ref{eqn:bistar}) as,
\begin{eqnarray}\label{eqn:uistar}
u_i^*=\left [1-\left ( \frac{(N-1)\beta_i}{\sum_{j=1}^N\beta_j}\right )\right ]^+,
\end{eqnarray}
where $x^+=\max(x,0)$. It is observed that the ownership
$u_i^*$ is non-zero if and only if (\ref{eqn:conditionnash}) is
satisfied, i.e., if the Nash equilibrium is feasible.
The condition in (\ref{eqn:conditionnash}) and the 
expression in (\ref{eqn:uistar}) have the following
interesting implications.
\begin{itemize}\itemsep -2pt
\item From (\ref{eqn:uistar}),
the ownership of contributors depend on
the $\beta_j$ of {\it all the contributors}. 
This is intuitively correct in a peer production
project like Wikipedia because contributions are
made by multiple users and the ownership held by 
a user will depend on the effort of all the users
that worked together in making contributions to the page.
\item The expression in (\ref{eqn:uistar}) indicates
that contributors who expend smaller effort have
larger ownership and those who expend larger effort
have low ownership, i.e., the fractional content ownership is a
decreasing function of the effort expended.
\item Asymptotically, i.e., as the number of contributors, $N$, becomes large,
the ownership, $u_i^*$ in (\ref{eqn:uistar}), can be written as
\begin{eqnarray}\label{eqn:uistarassym}
u_i*=\left ( 1-\frac{\beta_i}{E[\mbox{\boldmath $\beta$}]}\right )^+,
\end{eqnarray}
where $E[\mbox{\boldmath $\beta$}]\define\frac{1}{N}\sum_{j=1}^N\beta_j$,
is the {\it average effort} of all the users that make contributions
to the page. From (\ref{eqn:uistarassym}), only those contributors for whom
$\beta_i<E[\mbox{\boldmath $\beta$}]$, i.e., only those contributors
whose effort is below the average effort expended in posting content to
a page, end up with non-zero ownership. In other words, given the 
effort involved in making a contribution, and the ease in which others 
can overwrite one's contributions, only those who expend less effort 
in making their contributions than the average effort required,
end up with non-zero ownership.
\end{itemize}
\section{Empirical Validation with Data}\label{sec:results}
While the non-cooperative game theoretic models developed in 
Section \ref{sec:ncgame} are based on intuitive notions of 
ownership and effort, it is necessary to validate these with 
real data from contributions to Wikipedia articles. We require a set of Wikipedia 
articles with data on: (a) the content ``owned'' by 
contributors at each revision (which can be analogous
to the utility, $u_i^*$ in (\ref{eqn:uistar}) and (b) 
the cumulative effort exerted by each contributor 
(including all of his/her contributions) up to each 
revision, which can represent the effort, $\beta_i$, used in the
expressions in (\ref{eqn:finalexpressionxk}) and (\ref{eqn:uistar}). 
We use the data set from Arazy \emph{et al} \cite{oferjasist}, who 
explored automated techniques for estimated Wikipedia contributors' 
relative contributions. The data set in \cite{oferjasist} includes 
nine articles randomly selected from English Wikipedia. Each article 
was created over an average period of 3.5 years.
Section \ref{subsec:JASISTdata} presents the details
of the data set in \cite{oferjasist} and Section
\ref{subsec:verification} provides a validation of the 
same against the models developed in Section \ref{sec:ncgame}.

\subsection{Extracting data from Wikipedia Articles}\label{subsec:JASISTdata}
\begin{table*}[]
\begin{center}
\caption{\label{tab:oferdata} The list of articles for the data set in
\cite{oferjasist} and their attributes.}
\begin{tabular}{|l|l|l|r|r|r|}
\hline
Article title & Start Date & End Date & Duration & Edits & Unique\\
& (MM/DD/YYYY) & (MM/DD/YYYY) & (years) & & Editors\\
\hline
Aikodo \cite{aikido} & $11/29/2001$ & $06/13/2004$ & 2.5 & 72 & 62\\
\hline
Angel \cite{angel} & $11/30/2001$ & $12/09/2005$ & 4.0 & 341 & 277\\
\hline
Baryon \cite{baryon} & $02/25/2002$ & $ 08/25/2005$ & 3.5 & 73 & 62\\
\hline
Board Game \cite{boardgame}& $11/04/2001$ & $12/30/2004$ & 3.2& 220 & 155\\
\hline
Buckminster Fuller \cite{buckminster} & $12/13/2001$ & $07/14/2004$ & 2.6 & 65 & 55\\ 
\hline
Center for Disease & $10/16/2001$ & $03/05/2006$ & 4.4 & 65 & 58 \\
Control and Prevention \cite{center} & & & & &\\
\hline
Classical Mechanics \cite{classical} & $06/06/2002$ & $08/13/2006$ & 4.2 & 202 & 165\\
\hline
Dartmouth College \cite{dartmouth} & $10/01/2001$ & $08/26/2004$ & 2.9 & 70 & 55\\
\hline
Erin Brockovich \cite{erin} & $09/24/2001$ & $02/02/2006$ & 4.4 & 59 & 54\\
\hline
Total & & & 31.7& 1167 & 943\\
\hline
Average & & &3.5 & 129.7 & 104.8\\
\hline 
\end{tabular}
\end{center}
\end{table*}
The content ``owned'' by contributors at the end date of each article 
period was calculated using the method described in \cite{oferjasist}. 
A sentence was employed as the unit of analysis, and each full sentence 
was initially owned by the contributor who added it. As content on a 
wiki page evolves, a contributor may lose a sentence when more than 
50\% of that sentence was deleted or revised. A contributor making a 
major revision to a sentence can take ownership of that sentence. The 
algorithm tracks the evolution of content, recording the number of 
sentences owned by each contributor at each revision, until the study's 
end date. The effort exerted by each contributor was, too, based on the 
method and data set described in \cite{oferjasist}. Two research assistants 
worked independently to analyze every ``edit'' made to the 9 articles in the 
sample set and record: contributor's ID; the type of each ``edit'' 
to the wiki page (the categories used included: add new content, improve 
navigation, delete content, proofread, and add hyperlink); the scope of 
each edit (on a 5-point scale, from minor to major). For example, a 
particular edit might be categorized as major addition of new content. 
The two assessors reviewed the ``History'' section of articles (where 
Wikipedia keeps a log of all changes to a page), comparing subsequent 
versions. Once the assessors completed their independent work, and 
inter-rater agreement levels were calculated (yielding very high 
levels of agreement), the average of the two assessors was used in the 
analysis. Finally, the above data set was used to obtain the following
parameters on each Wikipedia article listed in Table \ref{tab:oferdata}:
\begin{itemize}
\item The number of exclusive contributors/users ($N$)
\item The total effort expended by the $i^{th}$ user ($1\leq i\leq N$), $s_i$
\item The number of edits made by the $i^{th}$ user ($1\leq i\leq N$), $e_i$
\item The number of sentences owned by the $i^{th}$ user ($1\leq i\leq N$), $p_i$.
\end{itemize}
The following subsection provides a detailed explanation
on how we use these parameters to verify the game theoretic
analysis described in Section \ref{sec:ncgame}.  
\subsection{Numerical verification of the analysis}\label{subsec:verification} 
Using the set of parameters obtained from the
pages in Table \ref{tab:oferdata}, listed in
Section \ref{subsec:JASISTdata}, we compute
the effort expended by user $i$ for unit
contribution, $\beta_i$, as
\begin{eqnarray}\label{eqn:betai}
\beta_i=\frac{s_i}{e_i}.
\end{eqnarray}
Using the $\beta_i$'s thus obtained, we use the expression in
(\ref{eqn:uistar}) to determine the estimated fractional
ownership on the Wikipedia page, that will be held by each contributor. We compare
this with the fraction $\frac{p_i}{\sum_{j=1}^N p_j}$ 
\begin{figure}
\centerline{\psfig{figure=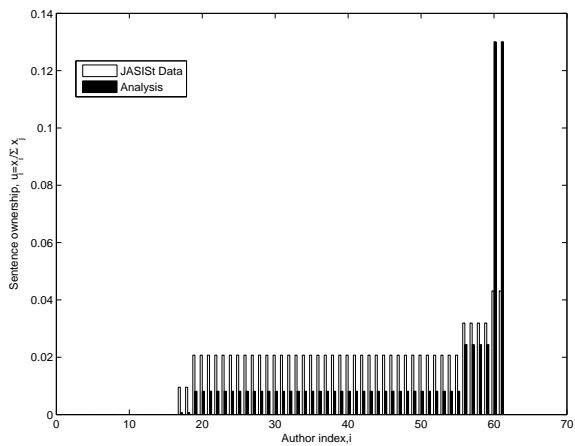,width=3in}}
\caption{\label{fig:aikido} Ownership of contributors obtained by the
game theoretic analysis presented in Section \ref{sec:ncgame} 
(described by the legend, ``Analysis'') and the 
data obtained from Wikipedia pages according to the algorithm in
\cite{oferjasist} (described by the legend, ``JASIST data''), for
the page, ``Aikido''. Contributors/Authors are indexed according to 
the decreasing order of effort, $\beta_i$'s.}
\end{figure}
\begin{figure}
\centerline{\psfig{figure=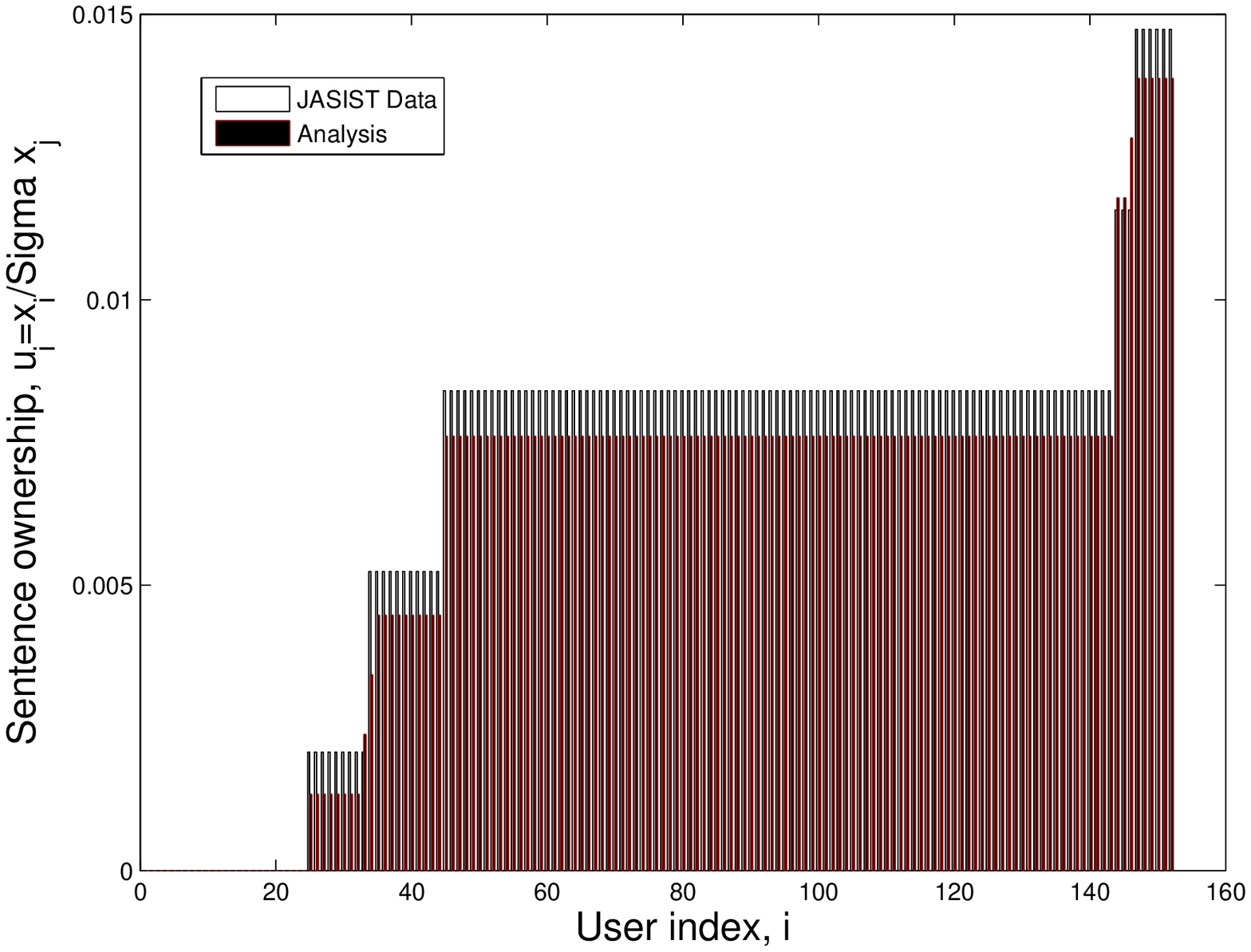,width=3in}}
\caption{\label{fig:board}Ownership of contributors obtained by the
game theoretic analysis presented in Section \ref{sec:ncgame} 
(described by the legend, ``Analysis'') and the 
data obtained from Wikipedia pages according to the algorithm in
\cite{oferjasist} (described by the legend, ``JASIST data''), for
the page, ``Board Game''. Contributors/Authors are indexed according 
to the decreasing order of  effort, $\beta_i$'s.}
\end{figure}
\begin{figure}
\centerline{\psfig{figure=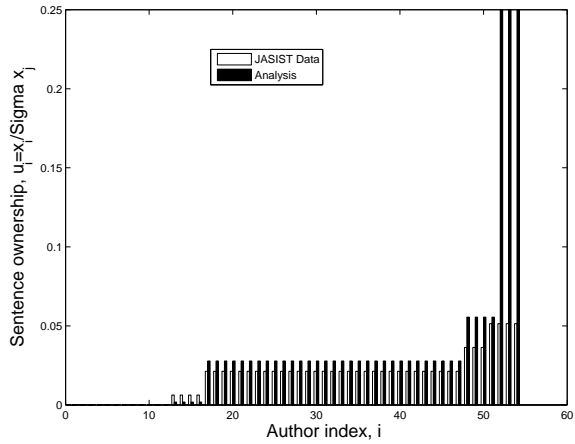,width=3in}}
\caption{\label{fig:erin} Ownership of contributors obtained by the
game theoretic analysis presented in Section \ref{sec:ncgame} 
(described by the legend, ``Analysis'') and the 
data obtained from Wikipedia pages according to the algorithm in
\cite{oferjasist} (described by the legend, ``JASIST data''), for
the page, ``Erin Brockovich''. Contributors are indexed according 
to the decreasing order of effort, $\beta_i$'s.}
\end{figure}
Figs. \ref{fig:aikido}, \ref{fig:board} and \ref{fig:erin} 
show the comparison between the ownership obtained 
according to the game theoretic analysis described 
in Section \ref{sec:ncgame} and that given by the 
data set in \cite{oferjasist} for the Wikipedia pages, 
``Aikido'',  ``Board Game'' and  ``Erin Brockovich'', 
respectively. For this first analysis, we anonymized 
the data set, indexing the users in the decreasing order of $\beta_i$’s. 
We find that the patterns in the empirical data and that of the 
game-theoretic model closely match one another.
In particular, the empirical data validates 
the following predictions made by the game theoretic
model in Section \ref{sec:ncgame}\footnote{These
trends were observed not only for the three articles shown in
Figs. \ref{fig:aikido}-\ref{fig:erin} but also for all the nine 
articles listed in Table \ref{tab:oferdata}. We 
show results for three articles here due to lack of space.}.
\begin{enumerate}[1.]
\item \textbf{Equivalence classes:} 
\begin{enumerate}
\item Let the users be classified into equivalence classes
according to their fractional ownership, i.e., all users
having equal fractional ownership in the Wikipedia page belongs to the same
equivalence class. It is observed that each page has five to six
equivalence classes. For instance, Aikido, has five
equivalence classes (Fig. \ref{fig:aikido}) and Board
game (Fig. \ref{fig:board}) and Erin Brockovich (Fig. \ref{fig:erin}), 
have six equivalence classes each.
\emph{Note that the number of equivalence classes obtained from
the data is the same as that predicted by the game theoretic analysis described
in Section \ref{sec:ncgame}}.
\item From (\ref{eqn:uistar}), $u_i^*=u_j^*$ if and only if
$\beta_i^*=\beta_j^*$. 
This indicates that the distribution of the data into 
number of equivalence classes applies not only to 
fractional ownership, but also to the effort expended by users.
In other words each Wiki page is expected to have five to six
categories of contributors/users. A more detailed analysis of 
the distribution suggests that the majority of users fall into 
the equivalence middle classes, while the classes on the extreme 
representing very low and very high levels of effort 
(and content ownership) comprise of relatively few users.
{\it While the above can be inferred from the data alone,
the game theoretic analysis provides a mathematical framework
that validates this observation.}
\end{enumerate}
\item \textbf{Non-zero ownership:} 
It is observed from (\ref{eqn:uistar}) that $u_i^*=0$
if and only if the condition in (\ref{eqn:conditionnash}) is violated.
The number of users in our sample data with zero ownership matches the 
number predicted by the game-theoretic model
thus providing validation for the condition 
(14) (at least for the Wikipedia pages included in our analysis).
{\it Again, it is observed that the relation between
the number of users with zero ownership and their corresponding $\beta_i$'s 
could have been inferred from the data alone, the game theoretic
analysis presented in Section \ref{sec:ncgame} provides a mathematical
framework to model this phenomenon.}
\end{enumerate}

\begin{figure}
\centerline{\psfig{figure=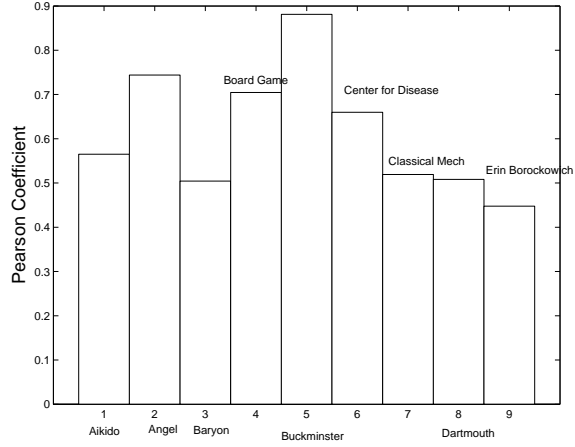,width=3in}}
\caption{\label{fig:pearson} Pearson correlation co-efficient between the values
of the fractional ownership, $u_i^*$, obtained from the data in 
\cite{oferjasist} and that obtained by the analysis in Section \ref{sec:ncgame}}
\end{figure}

After establishing that the general trend (i.e. anonymized 
data) for the empirical data and the model's predictions match one 
another, we perform a more detailed analysis where we pay 
attention to users' identities. That is, we organize both data 
sets, namely the fractional ownership data taken directly from 
\cite{oferjasist} and the ownership values our model in 
Section \ref{sec:ncgame} predicted, for each user. 
We then calculate the correlation between the two data sets, 
using the Pearson's correlation coefficient \cite{wolf}. 
The result of the analysis for the nine articles in our data 
set is presented in Fig. \ref{fig:pearson}. As could be seen 
from the figure, correlation coefficients range between 0.47 and 
0.88, representing moderate-high correlation. When combining the 
entire data from the nine articles into a single data set, the 
Pearson correlation was 0.65 (with a $p-$value, $p\approx 0.04$).
Therefore, we now proceed to verify if the discrepancies
in the values of the ownership obtained by the game
theoretic analysis and that obtained from the data
can be offset by establishing a linear fit that maps
the set of values obtained by analysis to the ones
obtained from the data. 
\begin{figure}
\centerline{\psfig{figure=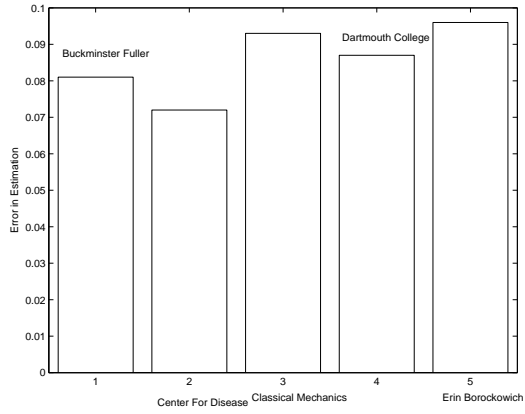,width=3in}}
\caption{\label{fig:error}Estimation error for the linear fit by the method of least
squares. The values of the fractional ownership, $u_i^*$, 
obtained for the pages, ``Aikido'', ``Angel'', ``Baryon'' and
``Board Game'' are used as training data for the linear fit.}
\end{figure}

Let $\mathbf{a}\define
\left [ 
\begin{array}{ccccc}
a_1 & a_2 & a_3 & \cdots & a_N
\end{array}
\right ]$ represent the ownership of the contributors obtained  
by the game theoretic analysis and let $\mathbf{d}\define
\left [ 
\begin{array}{ccccc}
d_1 & d_2 & d_3 & \cdots & d_N
\end{array}
\right ]$ represent the ownership of the contributors obtained
from the data as described in \cite{oferjasist}. For
each page, we fit
a function 
\begin{eqnarray}\label{eqn:linfit}
\begin{array}{cc}
\hat{d}_i=\rho a_i+\delta & 1\leq i\leq N,
\end{array}
\end{eqnarray}
where the parameters $\rho$ and $\delta$ are obtained
by the method of least squares \cite{meyerbook}. We 
use the values for the
pages ``Aikido'', ``Angel'', ``Baryon'' and ``Board Game'' as 
the training data to obtain $\rho$ and $\delta$. We then use
the values of $\rho$ and $\delta$ thus obtained to determine
$\hat{d}_i$ for the other five pages. We then compare $\hat{d}_i$
and $d_i$ and compute the estimation error for each page.
Fig. \ref{fig:error} shows the estimation error for
the remaining five pages. It is observed
that the error is between 7-9\%. The error for the
training set of data was found to be around 5\%. 
\emph{This indicates that the game theoretic analysis presented in Section 
\ref{sec:ncgame} models the contributor interactions 
in peer production projects
like Wikipedia accurately upto a linear scaling factor}.
\section{Trustworthy Collaboration and Vandalism}\label{sec:twcvand}
An important insight provided by our non-cooperative game 
model (and validated by our empirical analysis) is that only 
contributors with below-average effort levels are able to 
maintain fractional ownership on wiki pages. That is, by and 
large only the edits made by contributors who exert little 
effort survive the collaborative authoring process. In 
Section \ref{sec:intro}, we referred to two key concerns 
that are associated with trustworthy collaboration in peer-production 
projects: (a) a risk that non-experts will contribute content of 
low quality, and (b) a threat that malicious participants would 
vandalize Wikipedia pages. In spite of these serious concerns, 
the content on Wikipedia articles is generally of high quality 
and Wikipedia maintains the status as one of the most reliable 
sources of information on the web \cite{kittur}. 
How then, does Wikipedia maintain high-quality content in the 
face of threats of low-quality or malicious contributions? 
Our results can have important implications for investigation 
of trustworthy collaboration on Wikipedia (and more broadly, 
in peer-production projects). In the sections that follow, 
we provide two interpretations of our results that help 
explain how the threats highlighted above are mitigated.

\begin{enumerate}[1.]
\item \textbf{Trustworthiness/Quality of Wikipedia pages:} 
The first interpretation of the model and its empirical 
validation involves the concern of non-expert, low quality 
contributions eroding the trustworthiness of peer-produced product. 
This interpretation suggests that low effort is associated with 
greater likelihood of content survival due to a skill advantage: 
contributors who are experts in their field of contribution 
expend less effort, and their contributions are of higher quality
\cite{anthony}. 
Thus, the effort associated with contribution is inversely 
related with its quality and consequently with its likelihood 
of survival of subsequent editing. 
\item \textbf{Vandalism:} 
The second interpretation concerns the danger of 
vandalism activities reducing the trustworthiness of the 
peer-produced products. Since the underlying Wiki 
mechanisms allow any editor to easily revert the edits
of other contributors, the effort involved in vandalistic 
edits is higher than the effort of reverting such edits. 
Thus, high effort is associated with vandalism and 
relatively lower effort is linked to correction of 
vandalism. The game theoretic analysis presented in
Section \ref{sec:ncgame} predicts that the contributions
made by users expending large effort do not survive the edit
process and end up with zero ownership. Therefore, most 
vandalistic edits would not survive over time, as also observed
in \cite{kitturconflict}, \cite{suhchi2007}, \cite{stvtwi2008}.
\end{enumerate}

In summary, following on the intuition observed in 
\cite{kitturconflict}, \cite{suhchi2007}, \cite{stvtwi2008}, 
we modeled competition between players as a 
non-cooperative game, where a player's utility is 
associated with surviving fractional content owned, and 
cost is a function of effort exerted. 
Broader design implications emerging from this interpretation 
include the need to make version control mechanisms not only 
highly usable, but also highly open and egalitarian, and 
accessible to participants in a peer-production process. 
In addition, these insights suggest the importance of 
concurrent use of other quality control mechanisms, 
including user-designated alerts (where users are notified 
when changes are made to an article, or other part of the 
collaboratively-created product); watch lists (where users 
can track certain articles); and IP or user blocking in 
cases where repeated attacks from the same source are deemed 
to be acts of vandalism. The combination of these mechanisms 
make three important contributions to the trustworthiness 
of peer-production projects: first, their existence deter 
potential vandals; second, they reduce the costs of identifying 
and responding quickly to attacks; and third, they enable users 
to easily revert the consequences of vandalism .
\section{Conclusion}\label{sec:concl}
To better understand the success of peer production, 
we developed a non-cooperative game theoretic model 
of the creation of Wikipedia articles. 
The utility of a contributor was her relative 
ownership of the peer-produced product that 
survived a large number of iterations of collaborative 
editing. The work presented here contributes to better
understanding of the trustworthiness of peer-production by
\begin{itemize}
\item Solving the game and demonstrating the conditions 
under which a Nash equilibrium exists, showing that 
asymptotically only users with below average effort 
would maintain ownership
\item Empirically validating the model, demonstrating 
that only users with below average effort would maintain 
ownership, as well as showing editors' equivalence classes
\item Offering interpretations and implications for 
research on trustworthy peer-production (in terms of 
expertise and vandalism).  
\end{itemize}
To the best of our knowledge, this is the first modeling of 
user interactions on Wikipedia as a non-cooperative game. 
Our analysis points to the benefits of deploying multiple 
mechanisms which afford the combination of large-scale and 
low-effort quality control as way to ensure the trustworthiness 
of products created through web-based peer-production. 
Further research is needed to analyze the effectiveness 
of each of these mechanisms, and to address other aspects 
of peer production through game theoretic analysis.

\begin{center}
{\Large Acknowledgment}
\end{center}
This work was supported in part by a National Academies Keck Futures Initiative (NAKFI) grant.
\bibliographystyle{splncs03}
\bibliography{refs}
\end{document}